\documentclass[aps,prx,twocolumn,superscriptaddress]{revtex4-2}

\usepackage{amsmath,amssymb,graphicx}
\usepackage[final]{hyperref}
\usepackage{mathrsfs}
\usepackage{dcolumn}
\usepackage{bm}

\hypersetup{
colorlinks=true, 
linkcolor=blue, 
citecolor=blue, 
filecolor=magenta, 
urlcolor=blue
}

\usepackage{amsmath}
\usepackage{amsthm}
\usepackage{amssymb}
\usepackage{times}
\usepackage{braket}
\usepackage{ulem}

\usepackage{subcaption}
\usepackage[utf8]{inputenc}
\usepackage{hyperref}

\usepackage{graphicx, color}
    \graphicspath{{Figures/}}

\definecolor{chromeyellow}{rgb}{1.0, 0.65, 0.0}

\definecolor{light-blue}{rgb}{0.8,0.85,1}

\begin{document}

\title{Experimental Simulation of Loop Quantum Gravity on a Photonic Chip}

\author{Reinier van der Meer}
\affiliation{MESA+ Institute, University of Twente, P.O. box 217, 7500AE Enschede, The Netherlands }

\author{Zichang Huang}
\affiliation{State Key Laboratory of Surface Physics, Department of Physics, Center for Field Theory and Particle Physics, and Institute for Nanoelectronic devices and Quantum computing, Fudan University, Shanghai 200433, China}
 \affiliation{Shanghai Qi Zhi Institute, Shanghai 200030, China}

\author{Malaquias Correa Anguita}
\affiliation{MESA+ Institute, University of Twente, P.O. box 217, 7500AE Enschede, The Netherlands }

\author{Dongxue Qu}
\affiliation{Department of Physics, Florida Atlantic University, 777 Glades Road, Boca Raton, FL 33431, USA}

\author{Peter Hooijschuur}
\affiliation{MESA+ Institute, University of Twente, P.O. box 217, 7500AE Enschede, The Netherlands }

\author{Hongguang Liu}
\affiliation{Institut f\"ur Quantengravitation, Universit\"at Erlangen-N\"urnberg, Staudtstr. 7/B2, 91058 Erlangen, Germany}

\author{Muxin Han}
\email{hanm@fau.edu}
\affiliation{Department of Physics, Florida Atlantic University, 777 Glades Road, Boca Raton, FL 33431, USA}
\affiliation{Institut f\"ur Quantengravitation, Universit\"at Erlangen-N\"urnberg, Staudtstr. 7/B2, 91058 Erlangen, Germany}

\author{Jelmer J. Renema}
\email{j.j.renema@utwente.nl}
\affiliation{MESA+ Institute, University of Twente, P.O. box 217, 7500AE Enschede, The Netherlands } 

\author{Lior Cohen}
\email{lior.cohen3@mail.huji.ac.il}
\affiliation{%
Department of Electrical, Computer and Energy Engineering, University of Colorado Boulder, Colorado 80309, USA. 
}

\date{
	\today
}

\begin{abstract}
The unification of general relativity and quantum theory is one of the fascinating problems of modern physics. One leading solution is Loop Quantum Gravity (LQG). Simulating LQG may be important for providing predictions which can then be tested experimentally. However, such complex quantum simulations cannot run efficiently on classical computers, and quantum computers or simulators are needed. Here, we experimentally demonstrate quantum simulations of spinfoam amplitudes of LQG on an integrated photonics quantum processor. We simulate a basic transition of LQG and show that the derived spinfoam vertex amplitude falls within $4\%$ error with respect to the theoretical prediction, despite experimental imperfections. We also discuss how to generalize the simulation for more complex transitions, in realistic experimental conditions, which will eventually lead to a quantum advantage demonstration as well as expand the toolbox to investigate LQG.
\end{abstract}

\maketitle

\begin{figure*}[tb!]
    \includegraphics[width=1\textwidth,keepaspectratio]{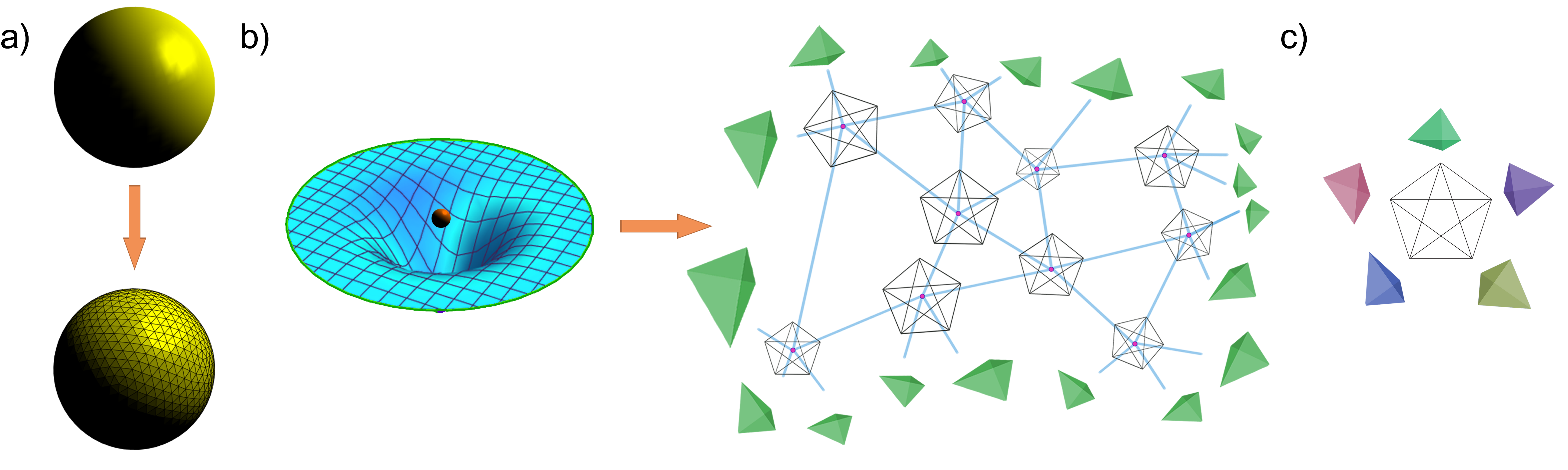}%
    \caption{\textbf{Concept of Loop Quantum Gravity.} a) A sphere is triangulated with a finite number of triangles, demonstrating triangulation of a simple geometry. In LQG the idea is to tile space-time with cells, called $4$-simplices. b) Illustration of discretization of continuous space-time into a complex, made of $4$-simplices. A line between two $4$-simplices indicates a shared tetrahedron, which is the 3D-hypersurface shared between two $4$-simplices. 
    c) A single $4$-simplex has 5 boundary tetrahedron states. In this work, a $4$-simplex is simulated with a linear photonic chip.  }
    \label{fig:fig1concept}
\end{figure*}

\section{Introduction.} 

We are currently in the era where the first devices which exhibit a quantum advantage have become available, i.e., devices which outperform a classical supercomputer at some well-defined computational task 
\cite{arute2019quantum,zhong2020quantum,wu2021strong,zhong2021phase}. Ultimately, quantum computing promises orders of magnitude speedup in solving problems of practical interest. Although some of these problems are connected to classical tasks \cite{shor1994algorithms,aaronson2011computational,hamilton2017gaussian}, many problems are concerned with solving the quantum dynamics of complex systems \cite{lloyd1996universal}. In the early 1980's, both Manin and Feynman independently conjectured that quantum dynamics cannot run efficiently on classical computers, but could be efficiently simulated on quantum systems \cite{Manin1980,feynman1985quantum,feynman1999simulating}. If the desired quantum dynamics are such that they can be simulated on a particular quantum system, implementation of a full-scale quantum computer may not be necessary. While universal, fault-tolerant quantum computers are possibly decades away, the timeline towards large-scale quantum simulators is potentially much shorter, making this an approach of interest for near-term applications of quantum computing. 

Integrated quantum photonics is one of the most promising platforms for quantum information processing protocols. Its small footprint and high stability make it a promising route for building large scale quantum systems \cite{moody2021roadmap}. Various quantum protocols have been suggested and demonstrated on photonic chips \cite{wang2020integrated}. In quantum communication, a few types of quantum key distribution have been realized on photonic chips \cite{ding2017high,sibson2017chip,lu2019chip}. In quantum computing, cluster states have been generated \cite{ciampini2016path}, also with error-protected logic qubits \cite{vigliar2021error}, for one-way quantum computing \cite{raussendorf2001one}. And in quantum simulation, photonic chips have been used for boson sampling \cite{aaronson2011computational,tillmann2013experimental,spring2013boson,wang2017high}, Gaussian boson sampling \cite{hamilton2017gaussian,zhong2021phase}, and quantum chemistry \cite{sparrow2018simulating, clements2018approximating}. Additionally, quantum simulations have been proposed for fundamental science \cite{preskill2018qftQuComp}, suitable for a photonic chip architecture \cite{Cohen:2020jlj}.

Loop Quantum Gravity (LQG) is a background-independent and non-perturbative approach to the theory of quantum gravity \cite{thiemann2008modern,han2007fundamental,ashtekar2004background}. As the LQG analog of the Feynman path integral for quantum gravity, the spinfoam amplitude is the transition amplitude for the evolution of the LQG quantum geometry state \cite{reisenberger1997sum,rovelli2014covariant,perez2013spin}. The spinfoam amplitude plays the central role in the covariant dynamics of LQG in 3+1 dimensions. The spinfoam amplitude is made of quantum gates that defines quantum transitions of quantum geometry states within Planck-scale volume regions (see Fig.~\ref{fig:fig1concept}) \cite{Cohen:2020jlj}. Matrix elements of these quantum gates are called vertex amplitudes. This feature of the spinfoam amplitude shares a similarity with systems in quantum computation and allows spinfoams to be demonstrated on a quantum simulator device \cite{Cohen:2020jlj,li2019quantum,mielczarek2019spin,Czelusta:2020ryq}.

Recently, a proposal was put forward for simulation of the spinfoam LQG on a linear-optical quantum simulator \cite{Cohen:2020jlj}. The key idea in this design is to map LQG quantum tetrahedron geometries (see Fig.~\ref{fig:fig1concept}\textcolor{blue}{(b)}) to optical modes, and to encode the spinfoam vertex amplitude in an optical quantum circuit as a chain of linear-optical unitary operations followed by post-selection.

In this work, we experimentally simulate a spinfoam vertex amplitude on a quantum photonic processor based on silicon nitride waveguides. The spinfoam vertex amplitudes are simulated on chip by encoding them in the unitary matrix $U$ that relates the input to output modes. 
The simulation not only encodes the vertex amplitude in a linear optical system, but also displays the semiclassical relation between the vertex amplitude and the geometry of a $4$-simplex. When scaling this experiment up to many modes, our experiment permits the simulation of spinfoam amplitudes with many vertices ($4$-simpleces), due to the inherent scalability of linear-optical quantum photonic processors and the generated path entanglement.



\section{Experimental setup}

\begin{figure*}[tb!]
    \includegraphics[width=1\textwidth,keepaspectratio]{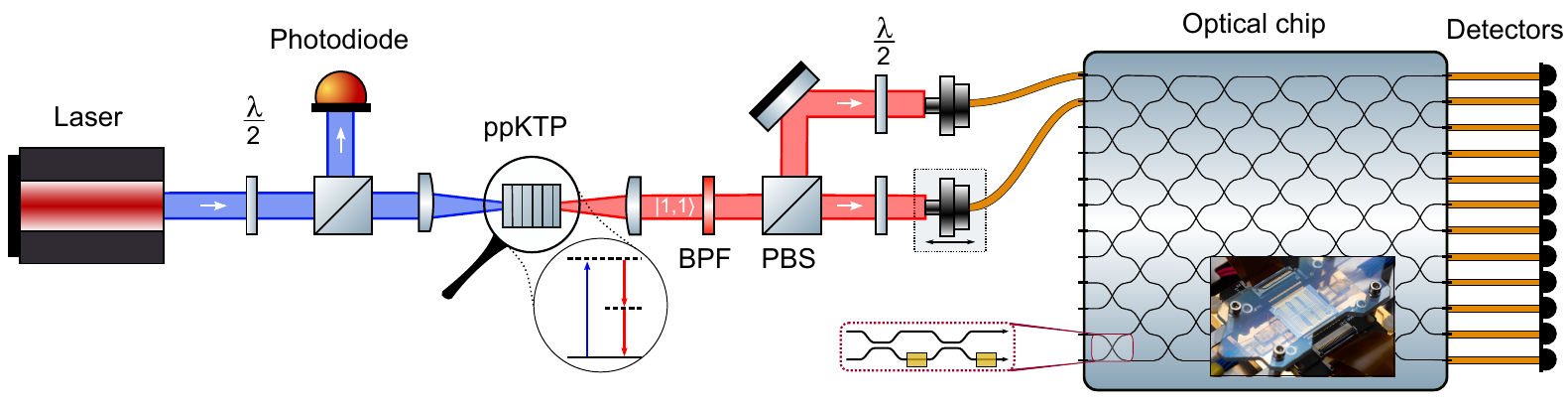}%
    \caption{\textbf{Setup.} A pulsed laser is used to generate pairs of photons in a ppKTP crystal. The generated photons have orthogonal polarizations and are separated by a polarizing beam splitter (PBS) and subsequently coupled into a polarization maintaining fibers which are connected to the optical network. After the optical network, the photons go through a single mode fiber to the single-photon detectors via a fiber polarization controller (not shown). The experiment requires proper indistinguishable photons. Hence, a $12\,$nm bandpass filter (BPF) is placed to remove spectral correlations between the photons. Furthermore, one of the fiber couplers is placed on a linear stage to guarantee temporal overlap of the photons. A beam sampler is used to monitor the power using a calibrated photo diode and the pump beam is filtered out after the ppKTP crystal (both not shown).}
    \label{fig:fig2Setup}
\end{figure*}

Our experimental setup (see Fig.~\ref{fig:fig2Setup}) is based on linear quantum optics, a non-universal platform for quantum simulations. In this model, bosonic interference between indistinguishable photons is used to process information encoded in the spatial degree of freedom of the photons. More specifically, we use a quantum photonic processor to implement the quantum simulation, and a single-photon source and single-photon detectors to test the quality of our implemented simulation.

The optical path of this experiment is as follows: heralded single-photon states (Fock states) are produced in a spontaneous parametric down-conversion (SPDC) \cite{evans_2010_Phys.Rev.Lett.} source on periodically-poled potassium titanyl phosphate (ppKTP). These single photons are then fed into our large-scale integrated quantum photonic processor, in which linear optical quantum interference occurs. Finally, the photons are measured at the output of the interferometer by superconducting nanowire single photon detectors (SNSPDs). 

The single-photon source consists of a Ti:Sapphire pulsed laser (Tsunami), producing pulses duration of $\Delta\tau\approx 100\,$fs centered at $\lambda=775\,$nm and with a width of $\Delta \lambda = 5.6\,$nm (full witdth at half maximum) at a repetition rate of $80\,$MHz. The laser is used to pump a $2\,$mm ppKTP non-linear crystal. Inside this crystal, a pump photon is spontaneously down-converted into a pair of single photons with degenerate spectra, centered at $\lambda=1550\,$nm. A spectral bandpass filter of $\Delta \lambda=12\,$nm (full width at half maximum) is used to remove any residual spectral correlations between the photons, and thus guaranteeing maximum indistinguishability. The photons are collected into polarization-maintaining single-mode optical fiber by fiber couplers, and fed into the chip at the desired input modes. The temporal overlap of the photons can be continuously tuned by a fibercoupler placed onto a motorized linear displacement stage. The Hong-Ou-Mandel (HOM) effect is used as a benchmarking tool to measure the degree of two-photon wave function overlap $x = \braket{\psi_{\text{photon}_1}|\psi_{\text{photon}_2}}$ according to $x^2 \geq V$, where $V$ is the visibility of the HOM dip \cite{Hong1987}. In our experiment we measure $x = 0.9899 \pm 0.0015$, which showcases the high quality of our source.

The photonic processor is a 12-channel integrated linear interferometer based on silicon nitride (Si$_3$N$_4$) waveguides, with an overall optical loss of $2.2 - 2.7\,$dB, corresponding to a transmission of 54-60\%, depending on the optical channel \cite{Taballione2021} \footnote{Note that the optical losses of the interferometer have been improved since this publication}. The linear network consists of an array of unit cells arranged in a square mesh architecture, whose geometry guarantees universality on the space of linear-optical transformations \cite{clements2016optimal}. Each unit cell of the interferometer corresponds to a Mach-Zehnder interferometer (MZI) between adjacent modes, and is tunable by the thermo-optic effect. For a full 12-mode transformation, the average amplitude fidelity is $F = 0.98$.

Detection of the photons after the interferometer is achieved using standard single-photon threshold (click) detectors. The output channels of the chip are connected via single-mode polarization-maintaining optical fiber to a bank of 12 superconducting nanowire single-photon detectors (SNSPDs), which are read out using conventional correlation electronics.

When performing measurements, the pump laser is operated at relatively low power ($\approx 5\,$mW), to avoid introducing cross-correlations between different experimental runs, since the dead time of the detctors is longer than the repetition rate of the source. At these power levels, with a photon-generation probability of $\sim$0.1\% per pulse, we achieve single photon rates of $19.6 \pm 0.16\,$kHz and two-photon coincidence rates of $2.56 \pm 0.11\,$kHz, where the errors correspond to Poissonian noise. 

\section{Results}
We implement a non-unitary $4\times8$ quantum gate of two input qubits to three output qubits corresponding to a specific $4$-simplex  \cite{Cohen:2020jlj}. The four-simplex is the analog of a vertex amplitude in a Feynman diagram. The vertex amplitude in QED is not unitary, although the full transition amplitude is. In LQG, a $4$-simplex quantum gate converts boundary states into spinfoam amplitudes (complex numbers). This is analogous to process amplitudes in quantum mechanics, which is equal to the scalar product of the input states and the output states separated by the process operator. The boundary of a $4$-simplex is made by five quantum tetrahedra (see Fig.~\ref{fig:fig1concept}\textcolor{blue}{(c)}). This gate is unitarized to a $12\times12$ transformation \cite{fiedler2009suborthogonality} and programmed on our processor \cite{clements2016optimal, Taballione2021}. 

To ascertain the quality with which we have done so, we must characterize the fidelity of our experimentally produced transition amplitude matrix with the target matrix. To measure this, we use the fact that a linear-optical transformation can be characterized with only one and two-photon measurements \cite{Laing2012,Dhand2016}. We measure $12$ single-photon transmission measurements and $21$ HOM dip measurements were carried out to infer amplitudes and phases of the unitary transformation respectively. We run the experiment for $60$ seconds per one-photon measurement, and $25$ minutes per HOM dip measurement. 

\begin{figure}
    \centering
    \includegraphics[width=0.5\textwidth,keepaspectratio]{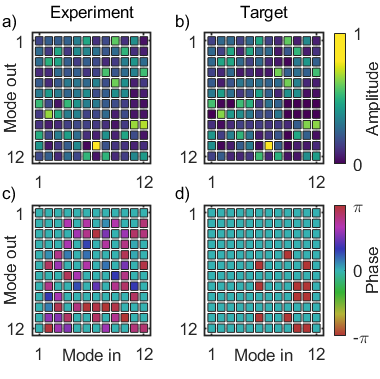}
    \caption{\textbf{Reconstructed Matrix.} The top row shows the matrix amplitudes for the experimentally observed matrix (a) and for the target matrix (b). The second row shows the matrix element's phases for the experiment (c) and target (d), respectively.}
    \label{fig:figResults}
 \end{figure}

Figure \ref{fig:figResults} shows the results of the matrix characterization of the experimentally measured matrix $U_\mathrm{\rm exp}$ against the target (theory) matrix $U_\mathrm{\rm th}$. In Figs.~\ref{fig:figResults}\textcolor{blue}{(a)} and  \textcolor{blue}{(b)} the normalized amplitudes of the matrix elements $|U_{i,j}|$ are plotted using a color scheme, for the experiment and theory matrices respectively. Each one of the colored blocks represents a matrix element of the unitary transformation that relates specific inputs and outputs of the photonic processor. A good agreement can be seen between the experiment and theory, as evidenced by the high amplitude fidelity $F = \frac{1}{N} \text{Tr}(|U^{\dagger}_{\text{target}}||U_{\text{\rm exp}}|) = 0.878$ for the entire $12\times12$ matrix and $F = 0.894$ for the $8\times4$ submatrix of interests corresponding to the $4$-simplex. 
In Figs.~\ref{fig:figResults}\textcolor{blue}{(c)} and \textcolor{blue}{(d)} the phases of the matrix elements $|U_{i,j}|$ are plotted using a color scheme, for the experiment and target matrices respectively. Each one of the colored blocks represents a matrix element of the unitary transformation that relates specific inputs and outputs of the photonic processor.

\begin{figure*}[htbp]
\centering
\includegraphics[width=1.7\columnwidth,keepaspectratio]{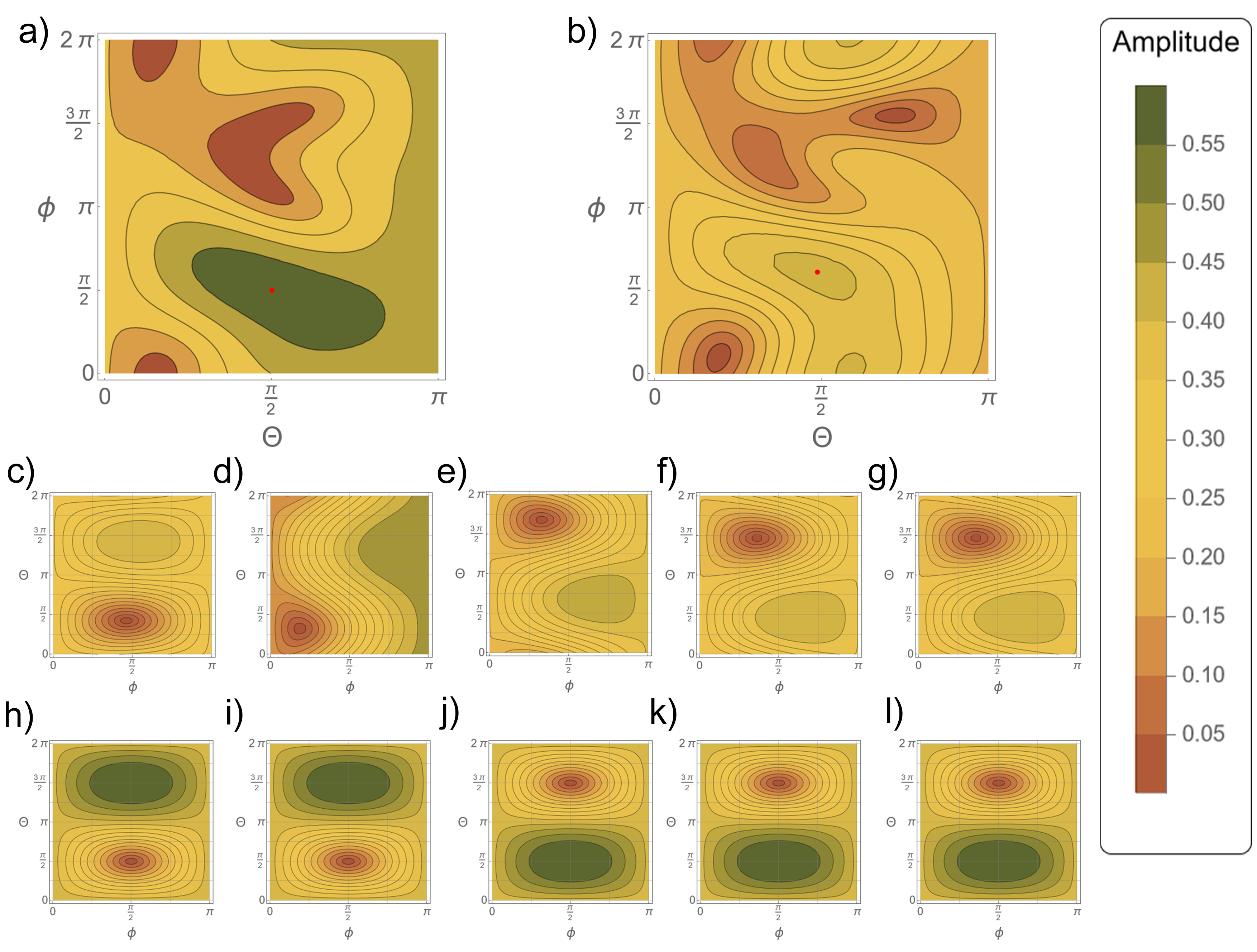}
\caption{\textbf{Contour Plots} The contour plots of the absolute values of the vertex amplitude given by different boundary states. $\theta$ and $\phi$ parameterize the boundary quantum tetrahedra states. (a) is calculated based on the experimental matrix and (b) by the theoretical matrix. The shapes of the contours in (a) and (b) share some common features, e.g., the Y-shape valley in the top half and the trough in the bottom-left corner. The maximum values in both plots are marked by red dots. The positions where these maximum values appear are nearly the same in both plots. (c)-(l) show the contour plot where 4 tetrahedra are regular (i.e. $(\theta,\phi)=(\pi/2,\pi/2)$ or $(\theta,\phi)=(\pi/2,3\pi/2)$) and we change $(\theta,\phi)$ of the fifth tetrahedron. (c)-(g) are calculated based on the experimental matrix and (h)-(l) by the theoretical matrix 
}\label{fig:Dis1}
\end{figure*}

To assess the quality of our implemented 4-symplex simulation from an LQG perspective, we choose a class of specific boundary states where all the face spins are $1/2$ and the quantum tetrahedra are independent of each other. Thus, these boundary states are tensor products of five single-qubit states \cite{li2019quantum}. 
Now, to estimate the degree of similarity between theory and experiment we compare the spinfoam amplitudes, computed with the chosen boundary states and the experimental and theoretical quantum gates.
The $4$-simplex is projected to a theoretical amplitude function, $A_{\rm th}(\theta_1,\phi_1\cdots\theta_5,\phi_5)$, depending on five pairs of inclination angle $\theta_i$ and azimuth angle $\phi_i$ on the Bloch spheres of the single-qubit states of the boundary state. 
Similarly, the  experimental $8\times4$ reduced matrix defines an amplitude function, $A_{\rm exp}(\theta_1,\phi_1\cdots\theta_5,\phi_5)$. 
We use these functions to compute spinfoam amplitudes for three types of boundary setups.
In the first setup, we let all the inclination angles and azimuth angles equal to $\theta$ and $\phi$. 
The contour plots of the experimental and theoretical amplitude functions are presented in Figs.~\ref{fig:Dis1}\textcolor{blue}{(a)} and \textcolor{blue}{(b)}. The shared features of the two contour plots indicate the agreement between theory and experiment, e.g., there is a "Y"-shape valley in the upper part of both plots, the positions of the peak in both plots are nearly the same. 
The second setup sets four out of five quantum tetrahedra as regular quantum tetrahedra, i.e., their angle pairs are either $(\pi/2,\pi/2)$ or $(\pi/2,3\pi/2)$ depending on the orientation of the $4$-simplex. In this setup, $A_{\rm th}$ and $A_{\rm exp}$ are functions depending on the $\theta$ and $\phi$ of the fifth tetrahedron.  Figures~\ref{fig:Dis1}\textcolor{blue}{(h)} to \textcolor{blue}{(l)} and \textcolor{blue}{(c)} to \textcolor{blue}{(g)} are the contour plots of $|A_{\rm th}|$ and $|A_{\rm exp}|$ given by varying a different quantum tetrahedron. 
In the third setup, the five boundary quantum tetrahedra are random states given by circular unitary distribution.
The expectation value of $A_{\rm th}$ is $0.0196+\mathrm{i} 0.000146$, and the expectation value of $A_{\rm exp}$ is $0.0204+\mathrm{i} 0.0000521$, which results in a percentage difference around $4.10\%$.

Furthermore, by LQG theory and given our boundary state class, $A_{\rm th}(\theta_1,\phi_1\cdots\theta_5,\phi_5)$ is invariant under the transformation of swaping between label $1$ and label $2$, and the permutations of labels $3$, $4$, and $5$. 
Fig.~\ref{fig:Dis1} \textcolor{blue}{(h)} to \textcolor{blue}{(l)} show this symmetry clearly. 
Fig.~\ref{fig:Dis1} \textcolor{blue}{(c)} to \textcolor{blue}{(g)}, except the large error in Fig.~\ref{fig:Dis1}\textcolor{blue}{(d)}, show that our $A_{\rm exp}(\theta_1,\phi_1\cdots\theta_5,\phi_5)$ approximately reproduces this symmetry.


\section{Discussion}
Both the fidelity $F$ and the comparison between the amplitude functions, $A_{\rm th}$ and $A_{\rm exp}$, show agreement between experiment and theory expectations with tolerable errors. 
When we set the boundary states as random states, the small difference between the theoretical and experimental expectation values supports this overall agreement from a different point of view, showing that the experimental operation of the chip carries physical information about the LQG of one vertex. 
In addition, the contour plots of Fig. \ref{fig:Dis1} show that one of the most important physical information is captured by our experimental matrix. 
In LQG, the saddle points where the amplitude reaches its extremum values carries the geometry information which is crucial to link the quantum theory to the gravity theory.
In both Fig.~\ref{fig:Dis1}\textcolor{blue}{(b)} and \textcolor{blue}{(c)}, the maximum value of the $|A_{\rm exp}|$ appears around $(\theta,\phi)=(\pi/2,\pi/2)$. 
Although with some error, Fig.~\ref{fig:Dis1}\textcolor{blue}{(d)} to \textcolor{blue}{(g)} also show a tendency that the maximum value would appears around $(\theta,\phi)=(\pi/2,\pi/2)$ or $(\theta,\phi)=(\pi/2,3\pi/2)$. 
In LQG \cite{li2019quantum}, the expectation values of the geometric operators on a qubit state with $(\theta,\phi)=(\pi/2,\pi/2)$ or with $(\theta,\phi)=(\pi/2,3\pi/2)$ indicate the quantum tetrahedron can be interpreted as a regular tetrahedron. 
This means that our experimental results show a trend that the most possible boundary configuration is given by the one that all boundary tetrahedra are regular.
In classical simplicial geometry, five regular tetrahedra make the boundary of a regular $4$-simplex. 
By this means, the quantum gate in our chip carries the proper physical information which makes the most possible boundary state as the one matches the classical simplicial geometry.

Aside with the aforementioned achievements, there are aspects yet to be improved. When setting all boundary tetrahedra to be regular, the resultant amplitude is $A_{\rm exp} = -0.363-\mathrm{i}0.183$. 
Compared to $A_{\rm th}=-0.287-\mathrm{i}0.497$, the percentage error of the absolute value of these quantities is around $29\%$. The peaks in Fig.~\ref{fig:Dis1}\textcolor{blue}{(d)} to \textcolor{blue}{(g)} deviate from the one giving regular tetrahedra. 
All deviations come from the errors in some elements of the matrix $U_{\rm exp}$. Decreasing these element-wise errors can be done by circuit optimization algorithms \cite{nam2018automated,fosel2021quantum} possibly by embedding the 12-mode chip in a larger chip, but this improvement is left for future research.

Scaling the experimental demonstration to larger number of vertices with more chips requires implementation of the vertices' connection. In general, this connection corresponds to a scalar product of two tetrahedra, applied by a projection of two quantum states. We highlight the question of implementing connections in an arbitrary encoding as a problem of interest. Since we implement the quantum states with spatial modes of single photons, the required projection involves photon-photon interaction, potentially using HOM two-photon interference \cite{pilnyak2019quantum}. 
Intuitively, the error is scaled linearly with the number of vertices. This is because the number of elements is linear in the number of vertices and although the transition amplitude is a nonlinear function of these elements since the error is small only the linear terms contribute and thus result in linear dependence. Further numerical investigation supports the above intuition. 

Another interesting topic is to analysis the computing complexity when we have a fixed triangulation with $N$ vertices. 
The recent interesting models in LQG include, e.g., a model of a black hole with $N=14$ \cite{Soltani:2021zmv} and a few models with $N=3,5,6$ on the semiclassical analysis \cite{Dona:2020tvv,Han:2021kll,Asante:2021zzh}. 
Denoting $C$ as the complexity of a gate representing one vertex with spin-$1/2$, the total complexity of the spin foam in this case is $C^N$.
The vertex amplitudes become more complicated, when spin is greater than $1/2$. 
In this case, two types of factors are introduced into the total complexity.
One is caused by the fact that each quantum tetrahedron $\Delta$ becomes a qudit whose complexity is bounded by $M_\Delta$. 
The other factor $J_f$ describes the effect caused by the summing over the spin of the internal triangle $f$.
Thus, in this case, the total complexity is bounded by $C^N\prod_\Delta M_\Delta \prod_f J_f$ where the $\prod_\Delta$ products over all the tetrahedra and $\prod_f$ is done over all the bulk triangles \cite{Cohen:2020jlj}.

The spinfoam model is a special case of tensor-network models \cite{Han:2018fmu}. The tensor-network models are generally made of quantum gates of qubits \cite{Orus:2013kga}, similar to the spinfoam amplitude that we study here. Thus, our experimental method should have wide applications to other tensor-network models. 

\section{Conclusion and outlook}

In conclusion, we have demonstrated that we can simulate a single $4$-simplex using a linear optical quantum system. We have shown that both by metrics current in the field of linear optics and by metrics oriented specifically towards the application of these systems in loop quantum gravity, we have faithfully implemented the properties of this $4$-simplex in linear optics. 




This demonstration, simulating a simple LQG transition with linear optics, is the first step towards full-scale simulations of loop quantum gravity, which is currently intractable with classical computers. 
The rapid growth in integrated photonic technology \cite{Taballione2021} is moving towards demonstration of quantum advantage. Using our method, this quantum advantage can be used as a tool for investigating LQG which is not available in today’s classical computing toolbox.

\begin{acknowledgements}

M.H. receives support from the National Science Foundation through grants PHY-1912278 and PHY-2207763. M.H. also acknowledges funding provided by the Alexander von Humboldt Foundation. R.v.d.M., P.H., M.C.A. and J.R. acknowledge funding from the Nederlandse Organisatie voor Wetenschappelijk Onderzoek (NWO) via QuantERA QUOMPLEX (Grant No. 680.91.037), and Veni (grant No. 15872). L.C. acknowledges funding from the National Science Foundation through Grant No. CCF-1838435.

\end{acknowledgements}

 \label{Experimental matrix}




\bibliographystyle{jhep}

\bibliography{LQGreferences.bib}

\end{document}